\documentclass[a4paper,11pt]{iopart}
\usepackage{graphicx}
\usepackage{cite}
\usepackage{color}
\usepackage{amsfonts}
\usepackage{srcltx}
\begin{document}

\title{Finite size effects in nonequilibrium wetting}
\author{A C Barato}
\address{The Abdus Salam International Centre for Theoretical Physics\\
    Trieste 34014, Italy}

\ead{acardoso@ictp.it}

\def\ex#1{\langle #1 \rangle}

\begin{abstract}
Models with a nonequilibrium wetting transition display a transition also in finite systems. This is different from nonequilibrium phase transitions into an absorbing state, where the stationary state is the absorbing one for any value of the control parameter in a finite system. In this paper, we study what kind of transition takes place in finite systems of nonequilibrium wetting models. By solving exactly a microscopic model with three and four sites and performing numerical simulations we show that the phase transition taking place in a finite system  is characterized by the average interface height performing a random walk at criticality and does not discriminate between the bounded-KPZ classes and the bounded-EW class. We also study the finite size scaling of the bKPZ universality classes, showing that it presents peculiar features in comparison with other universality classes of nonequilibrium phase transitions.
\end{abstract}

%=====================================================
\section{Introduction}
%=====================================================

It is well-known that phase transitions can only occur in systems with infinitely many degrees of freedom. For example, the Ising model in a finite volume does not display a transition, however using finite size scaling theory \cite{barber} one can obtain properties of the system in the thermodynamic limit by studying finite systems. A similar situation is observed for models in the so-called genuine nonequilibrium universality classes \cite{odor}, as is the case of direct percolation (DP) \cite{hinrichsenreview}.

On the other hand, a $d-$dimensional (with $d>2$) layer Ising model infinite in $d-1$ dimensions and finite with size $L$ in the remaining dimension displays a transition also for finite $L$. However, the transition for finite $L$ is different from the transition in the thermodynamic limit: In the first case we have the critical behavior of the Ising model in $(d-1)$-dimensions and in the second in $d$-dimensions. Similarly, nonequilibrium wetting finite systems also have a phase transition. Here the wetting transition is defined as an binding-unbinding transition controlled by a growth rate. The reason is that both phases are basically determined by the sign of the velocity of a free interface, giving a well-defined (but size-dependent) transition point, even for  finite systems. This circumstance has been pointed out some time ago by Mu\~noz \cite{munoz}. Nevertheless, the properties of this transition and its implications in finite size scaling were never studied in detail and this is the purpose of the present paper.

Nonequilibrium wetting is a very rich and interesting example of nonequilibrium critical phenomena. It basically corresponds to the study of the Kardar-Parisi-Zhang (KPZ) equation \cite{KPZ} added with a soft-wall potential ( see \cite{munoz04,santos04,barato10} for reviews). The KPZ equation defines a robust universality class of nonequilibrium growing free interfaces \cite{barabasi95,krug97}. It is a Langevin equation for the interface height $h({\bf x},t)$, where ${\bf x}$ gives the position on a $d$-dimensional interface and $t$ stands for time. The KPZ equation added with a term accounting for the presence of  a soft-wall was introduced by Tu et al. \cite{tu97} and reads
\begin{equation}
\frac{\partial h({\bf x},t)}{\partial t}= a-\frac{d}{dh}V(h)+\sigma\nabla^2 h({\bf x},t) +\lambda (\nabla h({\bf x},t))^2+\zeta({\bf x},t),
\label{eqbKPZ}
\end{equation}  
where $\zeta({\bf x},t)$ is a gaussian white noise and $V(h)$ is a soft-wall potential given by
\begin{equation}
V(h)= \exp(-h).
\end{equation}
The Laplacian term $\sigma\nabla^2 h({\bf x},t)$ is related to surface tension, $a$ is the average interface velocity at zero slope and the nonlinear term $\lambda (\nabla h({\bf x},t))^2$ is the lowest order term that breaks the up-down symmetry \cite{barabasi95,krug97}.

The wetting transition can be described as follows.  If $a$ is bigger than the critical value $a_c$ the interface will grow like a free KPZ interface, and the term $\exp(-h)$ becomes irrelevant after some transient. For $a<a_c$ the interface stays bounded to zero height and does not propagate. The order parameter for this transition is the ensemble average of $\exp(-h)$, which is zero for $a>a_c$ and non-zero for $a<a_c$. 

For $\lambda=0$ we have equilibrium wetting and equation (\ref{eqbKPZ}) becomes the Edwards-Wilkinson (EW) \cite{EW} equation  added with the soft-wall potential. This equation has been previously introduced by Lipowsky \cite{lipowsky85} in order to study dynamics in equilibrium wetting (see \cite{degennes85,dietrich86,bonn01,bonn09} for reviews on equilibrium wetting). For $\lambda\neq0$ we have nonequilibrium wetting and it turns out that the set of critical exponents characterizing the transition depends on the sign of $\lambda$. Therefore, we shall consider three universality classes: the bounded-EW class (bEW) and the bounded-KPZ classes (bKPZ+ and bKPZ--), where the sign refers to $\lambda$.  

By performing a Cole-Hopf transformation $n= \exp(-h)$ in equation (\ref{eqbKPZ}) the following multiplicative noise Langevin equation is obtained,
\begin{equation}
\frac{\partial n({\bf x},t)}{\partial t}= an(x,t)+n(x,t)^2+\sigma\nabla^2 n({\bf x},t)+n(x,t)\zeta({\bf x},t),
\label{eqMN1}
\end{equation}  
where $\lambda=-\sigma<0$. The above equation is called the multiplicative noise 1 (MN1) equation \cite{munoz04}. Performing the same transformation in equation (\ref{eqbKPZ}) but with $\lambda=\sigma>0$ the MN2 equation is obtained  \cite{munoz04}. Clearly  MN1 and bKPZ-- (MN2 and bKPZ+) are the same universality class. 

The most robust universality class of nonequilibrium phase transitions is the DP universality class \cite{hinrichsenreview}. The DP Langevin equation is very similar to the MN1 equation, the difference resides in the term multiplying the noise. In the MN1 equation it is proportional to $n(x,t)$, while in the DP equation it is proportional to $n(x,t)^{1/2}$. In spite of the similarity of both Langevin equations, the bKPZ-- (or MN1) and the DP universality classes are essentially different \cite{munoz}. Finite systems in the DP universality class have no phase transition, the stationary state is the absorbing one independently of the value of the control parameter. On the other hand, finite systems in the bKPZ universality classes still display a transition. This was clearly demonstrated by Mu\~noz \cite{munoz} with the solution of the corresponding one-variable Fokker-Planck equations (obtained by eliminating the Laplacian term) for the DP and the MN1 classes, which roughly corresponds to a zero-dimensional limit.   

In this paper we study the critical behavior of the phase transition taking place in a finite system and the finite size scaling of the bKPZ universality classes. We consider the microscopic model introduced in \cite{hinrichsen97}, which is a restricted solid on solid (RSOS) model with a hard-wall at zero height. We solve this model exactly for small systems with three and four sites, going beyond the zero-dimensional approximation from \cite{munoz}. As expected, we demonstrate that the critical exponents of the phase transition in finite systems are different from the critical exponents characterizing nonequilibrium wetting in the thermodynamic limit. Surprisingly, we find that the phase transition in finite-systems does not discriminate between the bEW, bKPZ+ and bKPZ-- universality classes. 

Moreover, we study the finite size scaling of bKPZ universality classes. We show that the survival of a transition also in a finite system leads to finite size scaling that is different from the finite size effects observed in the other genuine nonequilibrium universality classes \cite{odor}. In order to perform numerical simulations we use the single-step (SS) model with a moving wall \cite{ginelli03, kissinger05}. In this model the critical point is known exactly, leading to clean numerical results.

The organization of the paper is as follows. In the next section we introduce the models and discuss their phenomenology. In Sec. III we solve the RSOS model exactly for the three-site and four-site cases. In Sec. IV we analyze the finite size scaling of the bKPZ universality classes with numerical simulations. We conclude in Sec. V.      

%==========================================================================
\section{Models  definition and their critical behavior}
%==========================================================================

\begin{figure}
\begin{center}
\includegraphics[width=85mm]{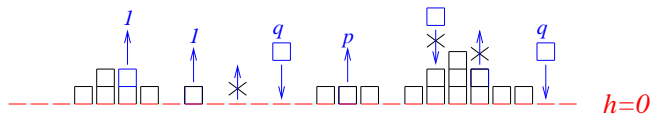}
\caption{Possible transitions in a certain interface configuration of the RSOS model with a wall. The crosses indicate that the transition cannot be carried out. } 
\label{transtionrates}
\end{center}
\end{figure}

First we define the microscopic realization of the bKPZ equation introduced by Hinrichsen et al.  \cite{hinrichsen97}. The model is defined on a one-dimensional discrete lattice of size $L$ with periodic boundary conditions. To each site $i$ of the lattice a random variable $h_i$ is attached. It can take any integer value and is interpreted as the interface height. The allowed configurations are constrained trough the RSOS restriction, $|h_{i\pm1}-h_i|\le 1$, which generates a surface tension. Usually, the initial condition is a flat interface at zero height, i.e. $h_i=0$ for all $i$. The model evolves by random-sequential updates and the possible transitions are: evaporation $h_i\to h_i-1$ and depositions $h_i\to h_{i}+1$. Only transitions that satisfy the RSOS restriction are carried out. The deposition rate (transition probability per unit of time) is $q$, and the evaporation rates are $p$ and $1$,  with $p$ being the evaporation rate at a plateau (see Fig. \ref{transtionrates}).  The hard-wall is introduced by forbidding evaporation events at zero height, in this way negative heights are not allowed.

%========================================%========================================%========================================
%========================================%========================================%========================================

\begin{figure}
\begin{center}
\includegraphics[width=75mm]{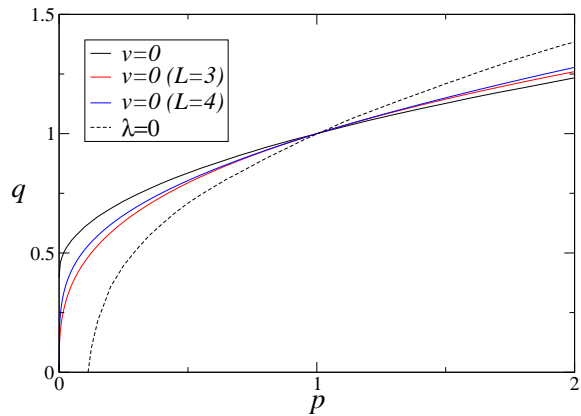}
\caption{ Phase diagram of the RSOS model. We show the lines $v=0$ and $\lambda=0$ which corresponds to $L\to \infty$ and are obtained from numerical simulations. Moreover the figure shows the exact critical lines for $L=3$ and $L=4$, which correspond to equation (\ref{eqcriticalL3}) and (\ref{eqcriticalL4}) respectively. } 
\label{phasediagram}
\end{center}
\end{figure}

In order to explain the wetting transition within this model we first consider a free interface (no wall at zero height). In this case the interface will propagate with a constant velocity after some transient. For a fixed value of $p$, if $q$ is bigger (smaller) than a certain value $q_c$, the interface propagates in the positive (negative) height direction. For $q=q_c$ the interface velocity is zero. Now consider the wall is present. For $q>q_c$ it has no influence because the interface moves in the positive height direction away from the wall. On the other hand, for $q<q_c$ the situation is completely different: since the interface cannot propagate in the negative height direction, it stays bounded to the wall. Therefore, at $q=q_c$ that is a wetting transition, with $q>q_c$ corresponding to the the wet (or moving) phase and $q<q_c$ corresponding to the bound phase. In Fig. \ref{phasediagram} we show the phase diagram of the model obtained from numerical simulations. We note that the critical line depends on the system size and we are denoting by $q_c$ the critical line in the limit $L\to\infty$. When considering the critical point of a finite system we write $q_c(L)$.
      
It can be shown that the parameter $\lambda$ in the KPZ equation can be obtained within a microscopic model  by calculating how the interface velocity vary with the interface slope \cite{barabasi95}. For the RSOS model the slope is introduced by changing the RSOS condition between sites $1$ and $L$ in the following way. In order to have an interface slope $m=M/L$ the RSOS condition between the boundary sites is changed to $h_L-h_1= M, M+1,M-1$. In Fig.  \ref{phasediagram} we also display the line $\lambda=0$ obtained from numerical simulations, where above (below) this line $\lambda<0$ ($\lambda>0$). Therefore, for $0<p<1$ the phase transition is in the bKPZ-- universality class and for $p>1$ in the bKPZ+ universality class. At $p=1$ and $q\le 1$ detailed balance holds (see \cite{hinrichsen97}) and we have equilibrium wetting, i.e. the phase transition is in the bEW universality class. 

%========================================%========================================%========================================
%========================================%========================================%========================================

The order parameter of the phase transition is the density of sites at zero height $\rho_0$, defined by 
\begin{equation}
\rho_0= \langle\sum_{i=1}^{L}\delta_{h_i,0}\rangle,
\end{equation}     
where the brackets indicate an ensemble average. In the stationary state it is zero for $q\ge q_c$ and positive for $q<q_c$. The critical exponent $\beta$ is related to how the stationary value of the order parameter goes to zero as criticality is approached from below, i.e.,
\begin{equation}
\rho_0^{st}\sim \Delta^\beta,
\label{defbeta}
\end{equation}
where $\rho_0^{st}$ is the stationary value of the density of sites at zero height, $\Delta=q_c-q$, and the above relation is valid for $\Delta$ small and positive. At the critical point the order parameter as a function of time goes to zero following a power law with exponent $\theta$, i.e.
\begin{equation}
\rho_0(t)\sim t^{-\theta}.
\label{deftheta}
\end{equation}
The critical exponents defined by the divergence of the temporal correlation length $\zeta_\parallel$ and the spatial correlation length $\zeta_\perp$ are 
\begin{equation}
\zeta_\parallel\sim \Delta^{-\nu_{\parallel}}\qquad\textrm{and}\qquad\zeta_\perp\sim \Delta^{-\nu_{\perp}},
\label{defnu}
\end{equation}
which are valid for $\Delta$ small and positive.

The KPZ universality class is related to the invariance of scale of the interface roughness \cite{barabasi95,krug97}. It turns out that the critical exponents of the bKPZ universality classes are associated with the scaling exponents of the KPZ universality class \cite{barato08}. Hence, it is known that, in one dimension, $\nu_\parallel= 3/2$ and $\nu_\perp= 1$ for the bKPZ universality classes and $\nu_\parallel= 2/3$ and $\nu_\perp= 4/3$ for the bEW class \cite{barato08}. From dimensional analysis one can see that $\theta= \beta/\nu_{\parallel}$, and, therefore, with the introduction of the wall only one new exponent arises (which can be $\theta$ or $\beta$). For the bEW this exponent can be determined exactly using the transfer matrix method  and its value is $\beta=1$ (or $\theta=  \beta/\nu_{\parallel}= 3/4$) \cite{hinrichsen97}. For the bKPZ universality classes no exact solution is known and the exponents are determined with numerical simulations. To our knowledge the best numerical values for these exponents were obtained in \cite{kissinger05} with time-dependent simulations of the SS model with a wall. The exponents are $\theta= 1.184(10)$ (or $\beta= \theta\nu_{\parallel}= 1.776(15)$) for the bKPZ-- universality class and $\theta= 0.228(5)$ (or $\beta=\theta\nu_{\parallel}=  0.342(8)$) for the bKPZ+ universality class. 

Let us now define the notation for the distance from criticality and the critical exponents of a finite system. The distance from criticality will be denoted by $\tilde{\Delta}=q_c(L)-q$. An important result that we assume to be valid for the bKPZ universality classes and has been verified numerically in \cite{barato08}, for the RSOS model, is that 
\begin{equation}
|\Delta-\tilde{\Delta}|\sim L^{-1/\nu_\perp}.
\label{deltadelta}
\end{equation}
The critical exponents for a finite system $\tilde{\beta}$ and $\tilde{\theta}$ are related to how the stationary value of $\rho_0$ approaches zero as $\tilde{\Delta}\to 0$ and how $\rho_0$ decays in time at the finite system critical point ($\tilde{\Delta}=0$), respectively. 

In this paper we also consider the SS model with a moving wall, which is more convenient to perform numerical simulations. In the following we describe this model briefly, for a full definition see \cite{ginelli03,kissinger05}. Similarly to the RSOS case this model is defined on a one-dimensional discrete lattice with periodic boundary conditions, however in the SS model the height difference between two neighbors is  $|h_{i\pm1}-h_i|=1$. The possible transitions are: deposition in a local minima with probability  $s$  and evaporation in a local maxima with probability $1-s$.  This model can be mapped onto the simple symmetric exclusion process and, therefore, its interface velocity is known and given by 
\begin{equation}
v(L)= (s-1/2)(1+1/L).
\label{eqvL}
\end{equation} 
The moving wall is implemented by increasing (or decreasing) the height of the wall by one unit after a fixed time interval $\Delta T$, which produces a wall velocity $v_W= 1/\Delta T$ ($v_W= -1/\Delta T$). If the wall velocity is equal to $v(L)$ we are at the critical point (of a finite system). We will consider the cases $s=1$ and $s=0$, which correspond to the bKPZ-- and bKPZ+ universality classes, respectively \cite{kissinger05}. All the critical exponents defined for the RSOS model are defined in the same way for the SS model, where the distances from criticality are given by $\Delta= |v_W-v(\infty)|$ and $\tilde{\Delta}= |v_W-v(L)|$. Note that from (\ref{eqvL}) we see that (\ref{deltadelta}) is exact for the SS model with a moving wall. The advantages of performing numerical simulations with this model come from the fact that the critical point is known exactly.    

In the next section we solve the RSOS model exactly for the cases $L=3$ and $L=4$ in order to make an exact analyses  of the stationary state of a finite system. Before ending this section we discuss the critical behavior of the RSOS model at $p=0$. This is a special case where the phase transition is in the DP universality class \cite{alon}. Comparing the transition at $p=0$ with the transition for $p\neq 0$ provides a simple explanation for the essential differences between the DP universality class and the bKPZ universality classes. At $p=0$ once a layer is filled it is not possible to evaporate a particle on it anymore. Hence, in a finite system along $p=0$ for any $q\neq 0$ the interface will have a positive velocity, since eventually a layer will be completed. Therefore, at $p=0$ a finite system has no phase transition. For $p\neq 0$ particles can be evaporated from complete layers and because of that a phase transition is still present in a finite system.

%==========================================================================
\section{Exact solution for $L=3$ and $L=4$}
%==========================================================================

Let us consider the RSOS model with $L=3$. We denote by $P(h,h',h'')$ the joint probability that site $1$ has height $h$, site $2$ has height $h'$ and site $3$ has height $h''$. For each height $h$ there are seven possible configurations that satisfy the RSOS restriction. Because of translation invariance some of these configurations have the same probability, reducing the number of variables we have to deal with. For example, $P(h,h,h+1)=P(h,h+1,h)=P(h+1,h,h)$. More precisely, we have three independent variables and they are: 
\begin{eqnarray}
x_h= P(h,h,h)\nonumber\\
y_h= P(h,h+1,h)\nonumber\\
z_h= P(h+1,h,h+1).
\end{eqnarray} 
Since there is a wall at zero height, $x_h=y_h=z_h=0$ for $h< 0$. 

The master equation for the three-site system reads
\begin{eqnarray}
\frac{d}{dt}x_h= y_h+q z_{h-1}-[q+p(1-\delta_{h,0})]x_h\nonumber\\
\frac{d}{dt}y_h= \frac{1}{3}\bigg(qx_h+2 z_{h}-(2q+1)y_h\bigg)\nonumber\\
\frac{d}{dt}z_h= \frac{1}{3}\bigg(2qy_h+p x_{h+1}-(q+2)z_h\bigg),
\label{masterL3}
\end{eqnarray}
where the above equations are valid for $h\ge 0$. As an example, consider the first equation. The term $y_h$ comes from the transition $(h,h+1,h)\to(h,h,h)$, which is an evaporation taking place with rate $1$, while the term $q z_{h-1}$ comes from the transition $(h,h-1,h)\to(h,h,h)$, which is a deposition taking place with rate $q$. On the other hand, the terms with minus sign come from the reversed transitions. The factor $(1-\delta_{h,0})$, also in the first equation, is related to the presence of the wall at zero height. The other two equations are derived in the same way.

Because of the RSOS restriction, the one-site probability distribution $P_h$ and the two-site probability distributions $P_{h,h'}$ are related by $P_h= P_{h,h}+P_{h,h+1}+P_{h,h-1}$. From the relation between the two-site probability distributions and the three-site probability distributions we obtain 
\begin{equation}
P_h= x_h+y_{h-1}+2y_h+2z_{h-1}+z_h.
\end{equation}
The order parameter in the present notation is the probability of being at height zero $P_0$. The stationary state solution of the master equation (\ref{masterL3}) gives
\begin{equation}
P_0^{st}= \frac{(1+q)^2}{p(1+3q+3q^2)}(p-q^3)
\end{equation}
where the above solution is valid below the critical line, given by
\begin{equation}
p= q^3.
\label{eqcriticalL3}
\end{equation}
Therefore, for $L=3$ the critical line is given by $p= q^3$ and $\tilde{\beta}=1$ for any value of $p$. By integrating equations (\ref{masterL3}) numerically at the critical point we obtain $\tilde{\theta}= 1/2$, also independent of the value of $p$. Hence, the critical behavior for this finite system is different from the critical behavior obtained in the limit $L\to\infty$. 

The case $L=3$ is a very peculiar situation because one cannot introduce an interface slope and therefore the notion of $\lambda$ is lost. In this sense it is not surprising that the critical behavior is the same, independent of the value of $p$. For this reason we proceed solving the system for $L=4$.  

In the case $L=4$ there are six independent variables, and they are
\begin{eqnarray}
x_h= P(h,h,h,h)\nonumber\\
y_h= P(h+1,h,h,h)\nonumber\\
z_h= P(h+1,h+1,h+1,h)\nonumber\\
u_h= P(h,h+1,h,h+1)\nonumber\\
v_h= P(h,h+1,h+1,h)\nonumber\\
w_h= P(h+1,h,h-1,h),
\label{eqfoursite}
\end{eqnarray} 
where $x_h=y_h=z_h=u_h=v_h=0$ for $h<0$ and $w_h=0$ for $h<1$.

The master equation for $L=4$ is given by
\begin{eqnarray}
 \frac{d}{dt}x_h= y_h+q z_{h-1}-[q+p(1-\delta_{h,0})]x_h\nonumber\\
 \frac{d}{dt}y_h= qx_h+u_h+2v_h+qw_h-[1+p(1-\delta_{h,0})+3q]y_h\nonumber\\
 \frac{d}{dt}z_h= px_{h+1}+qu_h+2qv_h+w_{h+1}-(p+2+2q)z_h\nonumber\\
 \frac{d}{dt}u_h= qy_h+pz_h-(1+q)u_h\nonumber\\
 \frac{d}{dt}v_h= 2qy_h+2z_h-(2+2q)v_h\nonumber\\
 \frac{d}{dt}w_h= qz_{h-1}+py_h-(q+1)w_h.
\label{masterL4}
\end{eqnarray}
The above equations are derived from the transition rules of the model, as we did for the three-site case. The one-site probability distribution as a function of the four-site probability distributions (\ref{eqfoursite}) reads
\begin{equation}
P_h= x_h+3y_h+y_{h-1}+z_h+3z_{h-1}+u_h+u_{h-1}+2v_h+2v_{h-1}+2w_h+w_{h-1}+w_{h+1}.
\end{equation}
The stationary state solution of equation (\ref{masterL4}) gives
\begin{equation}
P_0^{st}= \frac{2+3q(1+q)(2+q^2)+p(1+3q+3q^2)}{p(2+p)[(2+p)(1+4q+6q^2)+12q^3(q+1)}(2p+p^2-3q^4),
\end{equation}
which is valid below the critical line
\begin{equation}
p= -1+\sqrt{1+3q^4}.
\label{eqcriticalL4}
\end{equation}
Note that in both finite systems at $p=0$ we have $q_c=0$ (see equations (\ref{eqcriticalL3}) and (\ref{eqcriticalL4})), in agreement with the discussion in the previous section stating that at $p=0$, for a finite system, there is no phase transition. As in the three-site case, the critical exponents are again independent of $p$, and given by $\tilde{\beta}=1$ and $\tilde{\theta}=1/2$ (obtained from numerical integration of (\ref{masterL4}) at criticality). 

%========================================%============================
\begin{figure}
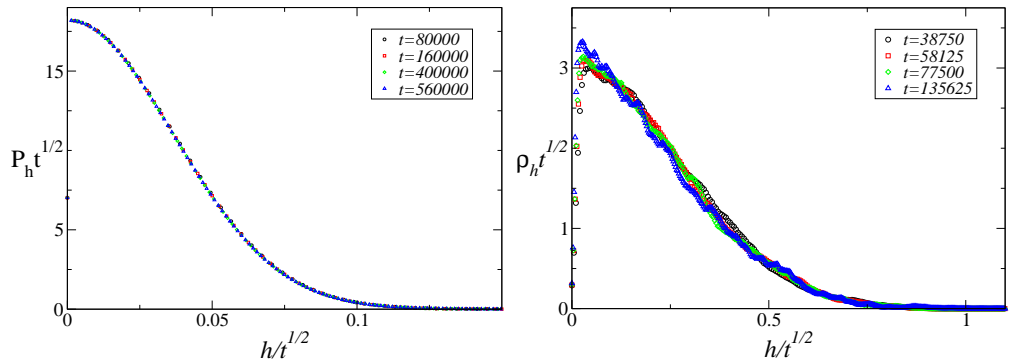

\begin{center}
\includegraphics[width=65mm]{Fig3a.eps}
\includegraphics[width=65mm]{Fig3b.eps}
\caption{ On the left panel we have the height profile obtained with numerical integration of the master equation (\ref{masterL4}) for $L=4$ for the RSOS model at the critical point $q=2$ and $p=6$. On the right panel: the height profile for the SS model at the critical point  in the bKPZ-- case ($s=1$) for L=32 obtained from numerical simulations ($5\times 10^5$ independent realizations). In both cases, the collapse of the data shows that the average height behaves like a random walk.} 
\label{figgaussian}
\end{center}
\end{figure} 
%========================================%============================

More precisely, by integrating the master equation numerically at criticality (for $L=3$ and $L=4$) we observe that the height profile is a gaussian scaling with $t^{1/2}$: the average interface height performs a random walk with a reflecting boundary (see Fig. \ref{figgaussian}). We also observed numerically (with Monte Carlo simulations)  that the critical behavior that we obtained here with the exact solution of the cases $L=3$ and $L=4$ holds for any finite system. As an example, in Fig. \ref{figgaussian} we show the gaussian height profile for $L=32$ obtained with numerical simulations of the SS model in the bKPZ-- case at the critical point. Surprisingly, the transition in a finite system is the same for all the three universality classes. We note that we can also write down the master equation for the case $L=5$ (there are 10 independent variables), which is considerably more cumbersome. However, we could not find an analytical expression for the critical line in this case.

We could consider a simpler case: a one site model, which would be a drifted random walk in the presence of a hard-wall (reflecting boundary). In this case, the drift is the control parameter and the critical point is at zero drift (where we have a reflecting boundary random walk). Therefore, already with this very simple approximation we have the critical behaviour of any finite system. Actually, this is the content of the one-component Langevin equation solved  in \cite{munoz}, which is obtained by eliminating the Laplacian from the MN1 equation. In the height variable it corresponds to a random walk in the presence of a soft-wall potential.

%==========================================================================
\section{Finite size scaling of the bKPZ universality classes}
%==========================================================================

The purpose of this section is to study the finite size scaling of the bKPZ universality classes. Since there is a transition also in a finite system,  the finite size scaling is different in comparison with the others genuine nonequilibrium universality classes \cite{odor}. For example, consider the DP universality class. In a finite system in the long time limit the system always goes to the absorbing state. The order parameter, in the active phase and near the critical point, as a function of time has initially a power law decay followed by a saturation at some value. If one considers times much bigger than the temporal correlation length (which depends on the system size) one sees a fast decay of the order parameter from the saturation value  to zero. Clearly for the bKPZ universality classes the situation is completely different. We expect to see a crossover (in off-critical and time-dependent simulations) from the infinite system critical behavior to the finite system critical behavior.      

%========================================%========================================%========================================
\begin{figure}
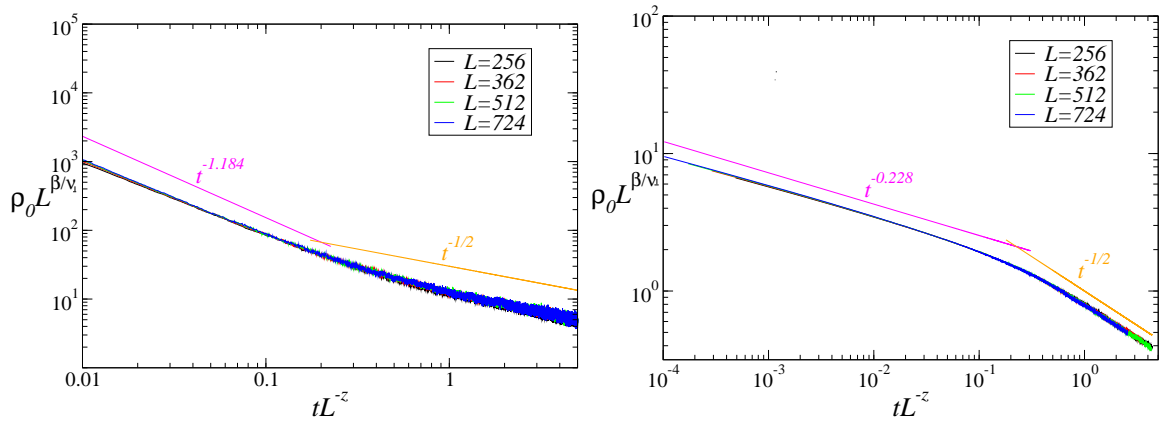

\begin{center}
\includegraphics[width=75mm]{Fig4a.eps}
\includegraphics[width=75mm]{Fig4b.eps}
\caption{Scaled Order parameter $\rho_0(t,L) L^{\beta/\nu_\perp}$ as function of the scaled time $t L^{-z}$ at the finite system critical point obtained from numerical simulations of the SS model. The crossover from $t^{-\theta}$ to $t^{1/2}$ is evident in the figure, for both universality classes.
}  
\label{figtemporal}
\end{center}
\end{figure} 
%========================================%========================================%========================================       

We show in Fig. \ref{figtemporal} time-dependent numerical simulations at the critical point (of the finite system) of the SS model for the bKPZ+ and the bKPZ-- universality classes. From the data collapse in Fig. \ref{figtemporal} we see that the order parameter at criticality has the scaling form
\begin{equation}
\rho_0(t,L)= L^{-\beta/\nu_\perp} f(t L^{-z}),
\label{scalingtemporal}
\end{equation}
where $f(x)$ is a scaling function that crosses over from $f(x)\sim x^{-\theta}$ for small values of $x$ to $f(x)\sim x^{-1/2}$ for bigger values of $x$. In other words, initially the order parameter decays with the exponent $\theta$ that characterizes the critical behavior for $L\to\infty$ and for times considerably larger than the temporal correlation length the decay exponent crosses over to the random walk exponent $1/2$.

\begin{figure}
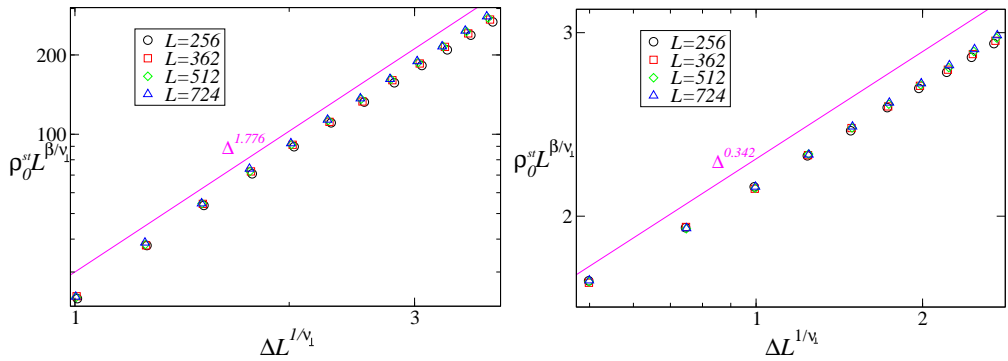

\begin{center}
\includegraphics[width=65mm]{Fig5a.eps}
\includegraphics[width=65mm]{Fig5b.eps}
\caption{The scaled order parameter in the stationary state $\rho_0^{st}(\Delta,L) L^{\beta/\nu_\perp}$ as a function of the scaled distance from criticality $\Delta L^{1/\nu_\perp}$ obtained from numerical simulations of the SS model.}
\label{figoff}
\end{center}
\end{figure}

For the off-critical simulations the usual scaling form is given by
\begin{equation}
\rho_0^{st}(\Delta,L)= L^{-\beta/\nu_\perp}g_i(\Delta L^{1/\nu_\perp}),
\label{scalingoffrho}
\end{equation}   
where $g_i(x)$ is a scaling function. The scaling function is expected to follow $g_{i}(x)\sim x^\beta$ for small $x$. In Fig. \ref{figoff} we plot $\rho_0^{st}(\Delta,L)L^{\beta/\nu_\perp}\times \Delta L^{1/\nu_\perp}$ for the bKPZ- and bKPZ+ universality classes obtained from numerical simulations of the SS model. We see a good data  collapse and the scaling function having the expected form.

 %========================================%========================================%========================================
%========================================%========================================%========================================       

\begin{figure}
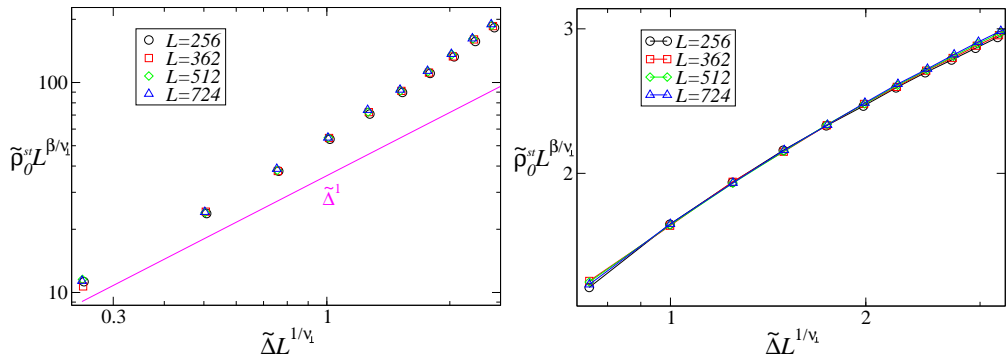

\begin{center}
\includegraphics[width=65mm]{Fig6a.eps}
\includegraphics[width=65mm]{Fig6b.eps}
\caption{The scaled order parameter in the stationary state, defined in (\ref{defrhotilde}), $\tilde\rho_0^{st}(\Delta,L) L^{\beta/\nu_\perp}$ as a function of the scaled distance from the finite system critical point $\tilde{\Delta} L^{1/\nu_\perp}$ obtained from numerical simulations of the SS model.}
\label{figoff2}
\end{center}
\end{figure}

Let us now analyse the order parameter in the stationary state as a function of the finite system distance from criticality. First we define the quantity
\begin{equation}
\tilde{\rho}_0^{st}(\tilde{\Delta},L)= \rho_0^{st}(\Delta,L).
\label{defrhotilde}
\end{equation}
From (\ref{deltadelta}) and (\ref{scalingoffrho}) we expect the following scaling form
\begin{equation}
\tilde{\rho}_0^{st}(\tilde{\Delta},L)= L^{-\beta/\nu_\perp}g_f(\tilde{\Delta} L^{1/\nu_\perp}),
\label{scalingoffrho2}
\end{equation} 
where the scaling function $g_f(x)$ should behave as $g_{f}(x)\sim x^{\tilde{\beta}}$ (with $\tilde{\beta}=1$) for small values of $x$. In Fig. \ref{figoff2} we plot this scaling function obtained from numerical simulations of the SS model with a wall. For the bKPZ-- case we see the scaling function approaching the expected linear form as we get closer to the critical point. For the bKPZ+ case we were not able to access regions close enough to criticality such that the form $g_f(x)\sim x$ sets in, considering the last two points for each system size we get an exponent of the order $0.57$, still far from $1$. However, one can clearly see an increasing curvature as $x$ gets smaller. 
 %========================================%========================================%========================================
%========================================%========================================%========================================       

We would like to stress three points. The first is that a similar situation for the off-critical finite size scaling has been considered in equilibrium in \cite{kastner00}, that is: a system with an exponent $\beta$ for $L\to\infty$ and a different exponent $\tilde{\beta}$ for finite $L$, independent of the system size. The same scaling forms (\ref{scalingoffrho}) and (\ref{scalingoffrho2}) were demonstrated to hold in \cite{kastner00}. Secondly, we think that previous numerical results of the critical exponent $\beta$ for the bKPZ-- class from off-critical simulations, as for example in \cite{hinrichsen97,barato08}, estimate a value smaller than $ 1.776(15)$ (obtained from time-dependent simulations in \cite{kissinger05}) because the finite-size effects were not took into account properly. The third point is that we also performed simulations with the RSOS model and the results are in agreement with the results for the SS model, however in the second case we get much better numerical results. 

%==========================================================================
\section{Conclusion}
%==========================================================================

As we cited in the introduction, it has been pointed out some time ago that the models in the bKPZ universality classes have a phase transition in finite systems \cite{munoz}, in opposition to the DP universality class. However, it was not clear what kind of transition takes place in a finite system. Here this issue was clarified: we showed that the phase transition in a finite system is characterized by a linear decay of the stationary value of the order parameter as criticality is approached and the average height of the interface performing a reflecting boundary random walk at the critical point. Moreover, we showed that the transition in a finite system does not differentiate between the bKPZ+, bKPZ-- and bEW universality classes. We have considered two microscopic realizations of the bKPZ equation: the RSOS and the SS model with a hard wall. However we expect this finite system phase transition to be a feature of the bKPZ universality classes.   

As another main result of this paper we have established the finite size scaling theory for the bKPZ universality classes, which is affected by the fact that finite systems also have a transition. For time-dependent simulations we showed that the scaling function (\ref{scalingtemporal}) has a crossover from the infinite system size decay exponent to the finite system size decay exponent. For the off-critical simulations we showed that the scaling function have different forms, depending on considering the distance from the infinite system critical point or the finite system critical point. In the first case the scaling function defined in (\ref{scalingoffrho}) follows $g_i(x)\sim x^\beta$ for small $x$ and in the second case the scaling function defined in (\ref{scalingoffrho2}) follows $g_f(x)\sim x^{\tilde{\beta}}$ for small $x$. 

Two remaining main open challenges in nonequilibrium wetting are the exact calculation of the new exponent, arising with the introduction of the wall, for the bKPZ classes and the experimental verification of the rich critical behavior that is predicted with the theory (see \cite{santos04,barato10} for a discussion). For the second challenge using the appropriate finite size scaling theory would be very important.

\noindent \textbf{Acknowledgment:}\\
I would like to thank Haye Hinrichsen for helpful discussions and carefully reading the manuscript. The Deutsche Forschungsgemeinschaft is gratefully acknowledged for partial financial support (HI 744/3-1).

\end{document}